# On the Performance of a Relay-Assisted Multi-Hop Asymmetric FSO/RF Communication System over Negative Exponential atmospheric turbulence with the effect of pointing error

Mohammad Ali Amirabadi

### Abstract

In this paper, a multi-user multi-hop hybrid FSO / RF system is presented. This structure is consisted of two main parts. The main motivation of presenting this structure is communication in long-range impassable links or some specific atmospheric conditions under which RF connection becomes easily disrupted. Although these effects could be mitigated by consuming more power or adding processing complexity, but a small user mobile phone cannot deserve more complexity or power supply. The fact that FSO and RF links are complementary of each other brings a new solution in mind; an access point that amplifies received signal via short-range RF link and forwards it to via long-range FSO link, could solve the mentioned problem. This scenario is exactly implemented at the first part of the proposed structure. At the second part, a multi-hop hybrid parallel FSO / RF link is implemented to connect source and destination Base Stations. It is the first time that in a multi-hop FSO / RF system, multi-user scheme, signal selection at each hop, known and un-known CSI in amplify and forward relaying, and saturate atmospheric turbulence with the effect of pointing error are considered. New expressions are derived in closed-form for Bit Error Rate (BER) and Outage Probability of the proposed structure and verified by MATLAB simulations. The proposed structure has advantages of FSO, RF, relay-assisted, and multi-user systems at the same time. Results indicate that it has low dependence on number of users and number of relays. Therefore, it is suitable for areas with varying population and long-range links. This structure offers independent performance without additional power consumption, processing latency, and complexity.

## I. INTRODUCTION

Free Space Optical (FSO) communication system often uses Intensity Modulation / Direct Detection (IM / DD) based on on-off keying. This scheme is simple and its detection threshold is changed based on atmospheric turbulence intensity; therefore, it is suitable for areas with frequent change of atmospheric turbulence intensity [1].

Due to advances of optoelectronic devices and reduction of their costs, FSO system has attracted many considerations during the last decades. It has large bandwidth and is highly secure, simple, cheap, and suitable for the last-mile backup application in 4G and 5G systems [2].

One of the main challenges toward FSO system is high sensitivity to atmospheric turbulences. In this effect, received signal intensity has random fluctuations [3]. Building vibration, which is caused by winds, brings misalignment of transreceiver. This effect is called pointing error, and can significantly degrade performance of FSO system. Various statistical distributions have been used in order to investigate the effects of atmospheric turbulences e.g., Exponential-Weibull [4], Generalized Malaga [5], Lognormal [6], Gamma-Gamma [7], and Negative Exponential [8]. Negative Exponential has high accompany with experimental results for saturate atmospheric turbulence.

Effects of weather conditions on FSO and RF links are not the same [9]; these two links are complementary of each other [10]. Accordingly, combination of FSO and RF links brings advantages such as reliability and accessibility at the same time.

Works done about the so-called hybrid FSO / RF system can be divided in three main categories. First category deals with single-hop parallel FSO / RF structure [9,11-17]; some works used multi-user scheme [9], but others used single-user scheme. Second category investigates performance of dual-hop structure [5, 7, 18-25]; some works used a direct backup link between source and destination [20], but others used either FSO or RF link at each hop. Last category deals with multi-hop structure. Multi-hop structure was previously investigated in FSO system [26-30], but in FSO / RF system it is considered as a new topic [31-34]. In these works, a single-user multi-hop hybrid parallel [32] or series [31,33,34] FSO / RF link is implemented and Outage Probability of the system is investigated. The main contributions and novelties of the proposed structure in this paper are assuming a different first hop structure, multi-user scheme, signal selection at each hop, amplify and forward plus demodulate and forward relaying, considering wide range of atmospheric turbulences from moderate to saturate with the effect of pointing error.

Many protocols have been used for relay-assisted FSO / RF system, e. g. decode and forward [19], amplify and forward [35], quantize and forward [20]. In amplify and forward, fixed gain amplification has less complexity but more power dissipation; thereby it is better to use only when Channel State Information (CSI) estimation is not possible. When CSI is known adaptive gain amplify and forward is a good choise; but generally speaking, demodulate and forward, due to its low power consumption is preferred [20].

In this paper, a multi-user multi-hop hybrid FSO / RF system is presented, which is consisted of two main parts. The main motivation of presenting this structure is that RF connection disrupts easily in long-range impassable links or under some specific atmospheric conditions. Although these effects could be mitigated by consuming more power or adding complex processing, but a small mobile phone battery cannot deserve more complexity or power consumption. The fact that FSO and RF links are complementary of each other brings a new solution in mind. An access point that amplifies received signal from user via short-range RF link and forwards it to the Base Station via long-range FSO link, could solve both of the mentioned problems.

At the first part of the proposed structure, users are connected to an access point via short-range RF link; between received RF signals, the maximum Signal to Noise Ratio (SNR) is selected, amplified and forwarded to source Base Station through a FSO link. At the second part, source and destination Base Stations are connected through a multi-hop parallel FSO / RF link with simultaneous data transmission at each hop. In the proposed multi-hop structure detect and forward protocol is used; because amplify and forward has high power consumption and is not efficient, also it amplifies noise as well as signal, therefore it is not affordable. Therefore, it is only recommended in long-range links where establishing the connection is more important than power consumption or complexity of the system.

Atmospheric turbulence of FSO link is modeled by Negative Exponential distribution in saturate regime. In order to get closer to the actual results, the effect of pointing error is also considered. Fading of RF link has Rayleigh distribution. For the first time, new closed-form expressions are derived for Bit Error Rate (BER) and Outage Probability of the proposed structure for Differential Binary Phase Shift Keying (DBPSK) modulation. Derived expressions are validated through MATLAB simulations.

The remainder of this paper is organized as follows: system model is expressed in Section II.

Performances of known CSI and unknown CSI schemes are investigated in sections III and IV, respectively. In section V simulation and analytical results are compared. Section VI is the conclusions of this study.

## II. System Model

Let $x_i$ be the transmitted RF signal from $i-th$ user; received signal at the first relay (access point) becomes as follows:

$$y_{1,i} = x_i h_{1,i} + e_{1,i}, \tag{1}$$

where $h_{1,i}$ is fading coefficient of RF channel between $i-th$ user and first relay, $e_{1,i}$ is additive white Gaussian noise (AWGN) with zero mean and $\sigma_{RF}^2$ variance. At each time slot, between received signals at the first relay, one with the highest SNR is selected. Let $y_1$ be the selected $y_{1,i}$; it is first converted to FSO signal by conversion efficiency of $\eta$, and then added by a unit amplitude bias to ensure that the FSO signal is not negative. The biased FSO signal is amplified and forwarded in the following form:

$$x_2 = G(1 + \eta y_1), \tag{2}$$

where $G$ is amplification gain. After removing biased component, the received signal at the second relay (source Base Station) becomes as follows:

$$y_2 = G\eta I_2 h_1 x + G\eta I_2 e_1 + e_2, \tag{3}$$

where $I_2$ is irradiance intensity of FSO channel, $e_2$ is AWGN with zero mean and $\sigma_{FSO}^2$ variance. Assuming that the transmitted signal have unit energy(i.e. $E[x^2] = 1$), instantaneous SNR at the second relay input becomes as follows [18]:

$$\gamma_{2^{nd}relay} = \frac{G^2 \eta^2 I_2^2 h_1^2}{G^2 \eta^2 I_2^2 \sigma_{RF}^2 + \sigma_{FSO}^2}. \tag{4}$$

In the case of known CSI for the first hop, the amplification gain could be adjusted according to the CSI; therefore, the output signal power is fixed. The relay gain is given by $G^2 = 1/(h_1^2 + \sigma_{RF}^2)$ [36]. Substituting $\gamma_1 = h_1^2/\sigma_{RF}^2$, $\gamma_2 = G^2\eta^2 I_2^2/\sigma_{FSO}^2$, and $G$, the end-to-end SNR at the second relay input becomes as follows [18]:

$$\gamma_{2^{nd}relay} = \frac{\gamma_1\gamma_2}{\gamma_1+\gamma_2+1}. \tag{5}$$

Regarding the fixed gain amplification, it could be chosen as $G^2 = 1/(C\sigma_{RF}^2)$, where $C$ is a desired constant parameter [36]. Substituting $\gamma_1 = h_1^2/\sigma_{RF}^2$, $\gamma_2 = G^2\eta^2 I_2^2/\sigma_{FSO}^2$, and $G$ into (4), the end-to-end SNR at the second relay input becomes as follows [5]:

$$\gamma_{2^{nd}relay} = \frac{\gamma_1\gamma_2}{C+\gamma_2}. \tag{6}$$

As mentioned in the introduction, amplify and forward relaying, due to its power consumption, is recommended only in situations that connection establishment is more important that power consumption or processing complexity. At the second part of the proposed structure, source and destination Base Stations are connected with multiple short hops without any connection problems. Therefore, demodulate and forward relaying, due to its simplicity and low power consumption is implemented at each relay station.

Received signal at the second relay ($y_2$) is demodulated, then regenerated, then modulated, then two copies of modulated signal are forwarded via parallel hybrid FSO / RF link. Assuming $M$ as the number of relays, at each hop, FSO channel has irradiance intensity of $I_j$; $j = 3,4, ..., M + 1$ and RF channel has fading coefficient of $h_j$. Received FSO and RF signals at each relay input are added by AWGN, with zero mean and $\sigma_{FSO}^2$ and $\sigma_{RF}^2$ variances. Signal with higher SNR is demodulated, then regenerated, then modulated and forwarded in both FSO and RF links. The same procedure is repeated at each hops.

The Cumulative Distribution Function (CDF) Rayleigh distribution is as follows [21]:

$$F_{\gamma_{RF}}(\gamma) = 1 - e^{-\frac{\gamma}{\bar{\gamma}_{RF}}}, \tag{7}$$

From Appendix A, the CDF of Negative Exponential atmospheric turbulence with the effect of pointing error is as follows:

$$F_{\gamma_{FSO}}(\gamma) = W \gamma^{\frac{\xi^2-1}{2}} G_{2,3}^{2,1}\left(\frac{\lambda}{A_0}\sqrt{\frac{\gamma}{\bar{\gamma}_{FSO}}}\;\middle|\; \begin{matrix} 1-\xi^2, 1 \\ 0, 2-\xi^2, -\xi^2 \end{matrix}\right). \tag{8}$$

Average electrical SNR in the FSO link is represented by $\bar{\gamma}_{FSO} = \eta^2 E(I^2)/\sigma_{FSO}^2 = \eta^2/\sigma_{FSO}^2$ [25].

Since users are separately located, their channels will experience independent fading. At each time slot, it is highly probable that at least one channel encounters favorable condition. The overall system performance will improve if the channel access is always granted to the best user, resulting in the so-called multi-user diversity gain. In this paper, the best user is assumed as one with the best link quality. Accordingly and considering [38], the CDF of the instantaneous SNR at the first relay input ($\gamma_1$) becomes as follows:

$$F_{\gamma_1}(\gamma) = \Pr\big(\max(\gamma_{1,1}, \gamma_{1,2}, \ldots, \gamma_{1,N}) \le \gamma\big) = \Pr\big(\gamma_{1,1} \le \gamma, \gamma_{1,2} \le \gamma, \ldots, \gamma_{1,N} \le \gamma\big), \tag{9}$$

where $\gamma_{1,i}$ is instantaneous SNR at the first relay input of $i-th$ path. Assuming independent identically distributed RF paths, and using (7), the CDF of instantaneous SNR at the first relay input becomes as follows [21]:

$$F_{\gamma_1}(\gamma) = \prod_{i=1}^{N} F_{\gamma_{1,i}}(\gamma) = \left(F_{\gamma_{1,i}}(\gamma)\right)^N = \left(1 - e^{-\frac{\gamma}{\bar{\gamma}_{RF}}}\right)^N. \tag{10}$$

Differentiating (10), the pdf of instantaneous SNR at the first relay input becomes as follows [21]:

$$f_{\gamma_1}(\gamma) = \frac{N}{\bar{\gamma}_{RF}} e^{-\frac{\gamma}{\bar{\gamma}_{RF}}} \left(1 - e^{-\frac{\gamma}{\bar{\gamma}_{RF}}}\right)^{N-1} = \sum_{k=0}^{N-1} \binom{N-1}{k} (-1)^k \frac{N}{\bar{\gamma}_{RF}} e^{-\frac{(k+1)\gamma}{\bar{\gamma}_{RF}}}. \tag{11}$$

In order to have better performance, between received FSO and RF signals at each relay, one with higher SNR is selected; accordingly, and considering [38], the CDF of the instantaneous SNR at $j-th$ relay input ($\gamma_j$), becomes as follows:

$$F_{\gamma_j}(\gamma) = \Pr\big(\max(\gamma_{\gamma_{FSO},j}, \gamma_{\gamma_{RF},j}) \leq \gamma\big) = \Pr\big(\gamma_{\gamma_{FSO},j} \leq \gamma, \gamma_{\gamma_{RF},j} \leq \gamma\big) = F_{\gamma_{FSO},j}(\gamma) F_{\gamma_{RF},j}(\gamma). \quad (12)$$

where the last equality is because FSO and RF links are assumed independent.

## III. Performance of known CSI scheme

Assuming high SNR approximation, (5) becomes equal to [18]:

$$\gamma_{2^{nd}relay} = \frac{\gamma_1 \gamma_2}{\gamma_1 + \gamma_2 + 1} \cong \min(\gamma_1, \gamma_2). \quad (13)$$

Therefore, the CDF of $\gamma_{2^{nd}relay}$ random variable becomes as [25]:

$$F_{\gamma_{2^{nd}relay}}(\gamma) = \Pr\big(\gamma_{2^{nd}relay} \leq \gamma\big) = 1 - \Pr\big(\min(\gamma_1, \gamma_2) \geq \gamma\big) = 1 - \Pr\big(\gamma_1 \geq \quad (14)$$
$$\gamma\big) \Pr\big(\gamma_2 \geq \gamma\big) = 1 - \big(1 - F_{\gamma_1}(\gamma)\big)\big(1 - F_{\gamma_2}(\gamma)\big).$$

Substituting (8) and (10) into (14), and using binomial expansion theorem, the CDF of $\gamma_{2^{nd}relay}$ in Negative Exponential atmospheric turbulence with the effect of pointing error becomes as follows:

$$F_{\gamma_{2^{nd}relay}}(\gamma) = 1 + \sum_{k=1}^{N} \binom{N}{k}(-1)^k e^{-\frac{k\gamma}{\bar{\gamma}_{RF}}} \left(1 - W\gamma^{\frac{\xi^2-1}{2}} G_{2,3}^{2,1}\left(\frac{\lambda}{A_0}\sqrt{\frac{\gamma}{\bar{\gamma}_{FSO}}}\bigg|_{0,2-\xi^2,-\xi^2}^{1-\xi^2,1}\right)\right). \quad (15)$$

### A. Outage Probability

In the proposed structure, outage occurs when instantaneous SNR at each relay input falls below a threshold level. Accordingly and considering [38], Outage Probability of the proposed structure becomes as follows:

$$P_{out}(\gamma_{th}) = \Pr\{(\gamma_1, \gamma_2, \ldots, \gamma_{M+1}) \leq \gamma_{th}\} = 1 - \Pr\{\gamma_1 \geq \gamma_{th}, \gamma_2 \geq \gamma_{th}, \ldots, \gamma_{M+1} \geq \gamma_{th}\} = 1 -$$
$$(1 - \Pr\{(\gamma_1, \gamma_2) \leq \gamma_{th}\}) \; (1 - \Pr\{\gamma_3 \leq \gamma_{th}\}) \ldots (1 - \Pr\{\gamma_{M+1} \leq \gamma_{th}\}) = 1 -$$
$$\left(1 - F_{\gamma_{2^{nd}relay}}(\gamma_{th})\right)\left(1 - F_{\gamma_3}(\gamma_{th})\right) \ldots \left(1 - F_{\gamma_{M+1}}(\gamma_{th})\right).$$

(16)

where the last equality is due to the fact that $P_{out}(\gamma_{th}) = F_\gamma(\gamma_{th})$. According to (12) and by substituting (15), (7) and (8) into (16), Outage Probability of the proposed structure for Negative Exponential atmospheric turbulence with the effect of pointing error becomes equal to:

$$P_{out}(\gamma_{th}) = 1 + \sum_{k=1}^{N} \binom{N}{k} (-1)^k e^{-\frac{k\gamma_{th}}{\bar{\gamma}_{RF}}} \left( W \gamma_{th}^{\frac{\xi^2-1}{2}} G_{2,3}^{2,1}\left(\frac{\lambda}{A_0}\sqrt{\frac{\gamma_{th}}{\bar{\gamma}_{FSO}}} \,\bigg|\, \begin{matrix} 1-\xi^2, 1 \\ 0, 2-\xi^2, -\xi^2 \end{matrix}\right)\right) \Bigg[ 1 -$$
$$\left(1 - e^{-\frac{\gamma_{th}}{\bar{\gamma}_{RF}}}\right) W \gamma_{th}^{\frac{\xi^2-1}{2}} G_{2,3}^{2,1}\left(\frac{\lambda}{A_0}\sqrt{\frac{\gamma_{th}}{\bar{\gamma}_{FSO}}} \,\bigg|\, \begin{matrix} 1-\xi^2, 1 \\ 0, 2-\xi^2, -\xi^2 \end{matrix}\right)\Bigg]^{M-1}.$$

(17)

Substituting binomial expansion of $\left[1 - \left(1 - e^{-\frac{\gamma_{th}}{\bar{\gamma}_{RF}}}\right) W \gamma_{th}^{\frac{\xi^2-1}{2}} G_{2,3}^{2,1}\left(\frac{\lambda}{A_0}\sqrt{\frac{\gamma_{th}}{\bar{\gamma}_{FSO}}} \,\big|\, \begin{matrix} 1-\xi^2, 1 \\ 0, 2-\xi^2, -\xi^2 \end{matrix}\right)\right]^{M-1}$

as $\sum_{t=0}^{M-1}\sum_{u=0}^{t} \binom{M-1}{t}\binom{t}{u} (-1)^{t+u} e^{-\frac{u\gamma_{th}}{\bar{\gamma}_{RF}}} \left( W \gamma_{th}^{\frac{\xi^2-1}{2}} G_{2,3}^{2,1}\left(\frac{\lambda}{A_0}\sqrt{\frac{\gamma_{th}}{\bar{\gamma}_{FSO}}} \,\big|\, \begin{matrix} 1-\xi^2, 1 \\ 0, 2-\xi^2, -\xi^2 \end{matrix}\right)\right)^t$, Outage

Probability of the proposed structure in Negative Exponential atmospheric turbulence with the effect of pointing error becomes equal to:

$$P_{out}(\gamma_{th}) = 1 +$$

(18)

$$\sum_{k=1}^{N}\sum_{t=0}^{M-1}\sum_{u=0}^{t} \Omega_2 \; e^{-\frac{(k+u)\gamma_{th}}{\bar{\gamma}_{RF}}} \left( W \gamma_{th}^{\frac{\xi^2-1}{2}} G_{2,3}^{2,1}\left(\frac{\lambda}{A_0}\sqrt{\frac{\gamma_{th}}{\bar{\gamma}_{FSO}}} \,\bigg|\, \begin{matrix} 1-\xi^2, 1 \\ 0, 2-\xi^2, -\xi^2 \end{matrix}\right)\right)^t \Bigg[ 1 -$$
$$W \gamma_{th}^{\frac{\xi^2-1}{2}} G_{2,3}^{2,1}\left(\frac{\lambda}{A_0}\sqrt{\frac{\gamma_{th}}{\bar{\gamma}_{FSO}}} \,\bigg|\, \begin{matrix} 1-\xi^2, 1 \\ 0, 2-\xi^2, -\xi^2 \end{matrix}\right)\Bigg],$$

where $\Omega_2 = \binom{N}{k}\binom{M-1}{t}\binom{t}{u}(-1)^{k+t+u}$.

## B. Bit Error Rate

In this paper DBPSK modulation is used, differential modulations such as DBPSK, are less sensitive to noise and interference, and because of the following reasons, their detection is optimum: no need for CSI, no need for feedback to adjust threshold, no effect on system throughput due to lack of pilot or training sequence, reduction in the effects of background noise at the receiver, reduction in the effects of pointing error [23]. The following expression can be used to calculate BER of DBPSK modulation [18]:

$$P_e = \frac{1}{2}\int_0^\infty e^{-\gamma} F_\gamma(\gamma)d\gamma = \frac{1}{2}\int_0^\infty e^{-\gamma} P_{out}(\gamma)d\gamma \tag{19}$$

Substituting (18) into (19), BER of DBPSK modulation in Negative Exponential atmospheric turbulence with the effect of pointing error becomes as follows:

$$P_e = \frac{1}{2}\int_0^\infty e^{-\gamma}\Bigg\{1 + \sum_{k=1}^{N}\sum_{t=0}^{M-1}\sum_{u=0}^{t}\Omega_2\, e^{-\frac{(k+u)\gamma}{\bar{\gamma}_{RF}}} \times \tag{20}$$

$$\left(W\gamma^{\frac{\xi^2-1}{2}}G_{2,3}^{2,1}\left(\frac{\lambda}{A_0}\sqrt{\frac{\gamma}{\bar{\gamma}_{FSO}}}\,\bigg|\,\begin{matrix}1-\xi^2,1\\0,2-\xi^2,-\xi^2\end{matrix}\right)\right)^t\left[1 - W\gamma^{\frac{\xi^2-1}{2}}G_{2,3}^{2,1}\left(\frac{\lambda}{A_0}\sqrt{\frac{\gamma}{\bar{\gamma}_{FSO}}}\,\bigg|\,\begin{matrix}1-\xi^2,1\\0,2-\xi^2,-\xi^2\end{matrix}\right)\right]\Bigg\}d\gamma.$$

When $t > 1$, the above integration is unsolvable; because there is no solution for the integration of multiplication of three Meijer-G and one exponential functions. Substituting (39) from Appendix B into (18), leads to a new exact form of Outage Probability of the proposed structure, substituting the result into (19) and using [37,Eq.07.34.21.0088.01], exact BER of DBPSK modulation in Negative Exponential atmospheric turbulence with the effect of pointing error becomes as follows:

$$P_e = \frac{1}{2}\Bigg\{ 1 + \sum_{k=1}^{N} \sum_{t=0}^{M-1} \sum_{u=0}^{t} \sum_{k_1=0}^{t} \sum_{k_2=0}^{k_1} \sum_{n=0}^{\infty} \Omega_2 \binom{t}{k_1} F_0^{t-k_1} E_n^{(k_1)} \Bigg[ \frac{\Gamma(1+H)}{\left(1+\frac{k+u}{\bar{\gamma}_{RF}}\right)^{1+H}} - $$

$$\frac{W 2^{-\xi^2}}{\sqrt{\pi}\left(1+\frac{k+u}{\bar{\gamma}_{RF}}\right)^H} G_{5,6}^{4,3}\left( \frac{\lambda^2}{4A_0^2 \bar{\gamma}_{FSO}\left(1+\frac{k+u}{\bar{\gamma}_{RF}}\right)} \Bigg| \begin{matrix} 1-H, \varrho_1 \\ \varrho_2 \end{matrix} \right)\Bigg]\Bigg\}. \tag{21}$$

where $\varrho_1 = \left(\frac{1-\xi^2}{2}, \frac{2-\xi^2}{2}, \frac{1}{2}, 1\right)$, $\varrho_2 = \left(0, \frac{1}{2}, \frac{2-\xi^2}{2}, \frac{3-\xi^2}{2}, -\frac{\xi^2}{2}, \frac{1-\xi^2}{2}\right)$.

## IV. Performance of unknown CSI scheme

Using (6), the CDF of end-to-end SNR at the second relay ($\gamma_{2^{nd}relay}$) becomes as follows [21]:

$$F_{\gamma_{2^{nd}relay}}(\gamma) = \Pr(\gamma_{FSO} \leq \gamma) = 1 - \Pr(\gamma_{FSO} \geq \gamma) = 1 - \Pr\left(\frac{\gamma_1 \gamma_2}{\gamma_2 + C} \geq \gamma\right). \tag{22}$$

After some mathematical simplifications, (22) becomes as follows [24]:

$$F_{\gamma_{2^{nd}relay}}(\gamma) = 1 - \int_0^\infty \Pr\left(\gamma_2 \geq \frac{\gamma C}{x} \Big| \gamma_1\right) f_{\gamma_1}(x+\gamma)dx. \tag{23}$$

Substituting (8) and (11) into (23), the CDF of $\gamma_{2^{nd}relay}$ in Negative Exponential atmospheric turbulence with the effect of pointing error becomes equal to [21]:

$$F_{\gamma_{2^{nd}relay}}(\gamma) = 1 - \sum_{k=0}^{N-1} \binom{N-1}{k} (-1)^k \frac{N}{\bar{\gamma}_{RF}} e^{-\frac{(k+1)\gamma}{\bar{\gamma}_{RF}}} \int_0^\infty e^{-\frac{(k+1)x}{\bar{\gamma}_{RF}}} \times \tag{24}$$

$$\left(1 - W\left(\frac{\gamma C}{x}\right)^{\frac{\xi^2-1}{2}} G_{2,3}^{2,1}\left(\frac{\lambda}{A_0}\sqrt{\frac{\gamma C}{x \bar{\gamma}_{FSO}}} \Bigg| \begin{matrix} 1-\xi^2, 1 \\ 0, 2-\xi^2, -\xi^2 \end{matrix} \right)\right) dx.$$

Substituting equivalent Meijer-G equivalent of $G_{2,3}^{2,1}\left(\frac{\lambda}{A_0}\sqrt{\frac{\gamma c}{x\bar{\gamma}_{FSO}}}\Big|\begin{matrix}1-\xi^2,1\\0,2-\xi^2,-\xi^2\end{matrix}\right)$

as $G_{3,2}^{1,2}\left(\frac{A_0}{\lambda}\sqrt{\frac{x\bar{\gamma}_{FSO}}{\gamma c}}\Big|\begin{matrix}1,\xi^2-1,\xi^2+1\\0,\xi^2\end{matrix}\right)$ [37, Eq.07.34.17.0012.01], and using [37, Eq. 07.34.21.0088.01] and [33, Eq.07.34.17.0012.01], the CDF of $\gamma_{2^{nd}relay}$ in Negative Exponential atmospheric turbulence with the effect of pointing error becomes equal to:

$$F_{\gamma_{2^{nd}relay}}(\gamma) = 1 - \sum_{k=0}^{N-1}\binom{N-1}{k}(-1)^k\frac{N}{k+1}e^{-\frac{(k+1)\gamma}{\bar{\gamma}_{RF}}}\Bigg(1-$$ (25)

$$W(\gamma c)^{\frac{\xi^2-1}{2}}\frac{2^{1-\xi^2}}{\sqrt{\pi}}\left(\frac{k+1}{\bar{\gamma}_{RF}}\right)^{\frac{1-\xi^2}{2}}G_{4,7}^{5,2}\left(\frac{\lambda^2c(k+1)\gamma}{4A_0^2\bar{\gamma}_{FSO}\bar{\gamma}_{RF}}\Big|\begin{matrix}\varphi_1\\\varphi_2\end{matrix}\right)\Bigg).$$

where $\varphi_1 = \left\{\frac{2-\xi^2}{2},\frac{1-\xi^2}{2},1,\frac{1}{2}\right\}$ and $\varphi_2 = \left\{\frac{3-\xi^2}{2},\frac{1}{2},0,\frac{3-\xi^2}{2},\frac{2-\xi^2}{2},\frac{1-\xi^2}{2},\frac{-\xi^2}{2}\right\}$.

## A. Outage Probability

Substituting (25), (7) and (8) into (16), Outage Probability of the proposed structure in Negative Exponential atmospheric turbulence with the effect of pointing error becomes equal to:

$$P_{out}\left(\gamma_{th}\right) = 1 - \sum_{k=0}^{N-1}\binom{N-1}{k}(-1)^k\frac{N}{k+1}e^{-\frac{(k+1)\gamma_{th}}{\bar{\gamma}_{RF}}}\Bigg(1-W(\gamma_{th}c)^{\frac{\xi^2-1}{2}}\frac{2^{1-\xi^2}}{\sqrt{\pi}}\left(\frac{k+1}{\bar{\gamma}_{RF}}\right)^{\frac{1-\xi^2}{2}}\times$$ (26)

$$G_{4,7}^{5,2}\left(\frac{\lambda^2c(k+1)\gamma_{th}}{4A_0^2\bar{\gamma}_{FSO}\bar{\gamma}_{RF}}\Big|\begin{matrix}\varphi_1\\\varphi_2\end{matrix}\right)\Bigg)\left[1-\left(1-e^{-\frac{\gamma_{th}}{\bar{\gamma}_{RF}}}\right)W\gamma_{th}^{\frac{\xi^2-1}{2}}G_{2,3}^{2,1}\left(\frac{\lambda}{A_0}\sqrt{\frac{\gamma_{th}}{\bar{\gamma}_{FSO}}}\Big|\begin{matrix}1-\xi^2,1\\0,2-\xi^2,-\xi^2\end{matrix}\right)\right]^{M-1}.$$

Substituting binomial expansion of $\left[1-\left(1-e^{-\frac{\gamma_{th}}{\bar{\gamma}_{RF}}}\right)W\gamma_{th}^{\frac{\xi^2-1}{2}}\times\right.$

$\left.G_{2,3}^{2,1}\left(\frac{\lambda}{A_0}\sqrt{\frac{\gamma_{th}}{\bar{\gamma}_{FSO}}}\Big|\begin{matrix}1-\xi^2,1\\0,2-\xi^2,-\xi^2\end{matrix}\right)\right]^{M-1}$ as $\sum_{t=0}^{M-1}\sum_{u=0}^{t}\binom{M-1}{t}\binom{t}{u}(-1)^{t+u}e^{-\frac{u\gamma_{th}}{\bar{\gamma}_{RF}}}\left(W\gamma_{th}^{\frac{\xi^2-1}{2}}\times\right.$

$G_{2,3}^{2,1}\left(\frac{\lambda}{A_0}\sqrt{\frac{\gamma_{th}}{\bar{\gamma}_{FSO}}}\Big|\begin{matrix}1-\xi^2,1\\0,2-\xi^2,-\xi^2\end{matrix}\right)\Big)^t$, Outage Probability of the proposed structure in Negative

Exponential atmospheric turbulence with the effect of poiting error becomes equal to:

$$P_{out}\left(\gamma_{th}\right)=1-\sum_{k=0}^{N-1}\sum_{t=0}^{M-1}\sum_{u=0}^{t}\Omega_4\,e^{-\frac{(k+u+1)\gamma_{th}}{\bar{\gamma}_{RF}}}\left(1-W(\gamma_{th}c)^{\frac{\xi^2-1}{2}}\frac{2^{1-\xi^2}}{\sqrt{\pi}}\left(\frac{k+1}{\bar{\gamma}_{RF}}\right)^{\frac{1-\xi^2}{2}}\times\right.$$

$$\left. G_{4,7}^{5,2}\left(\frac{\lambda^2 c(k+1)\gamma_{th}}{4A_0^2\bar{\gamma}_{FSO}\bar{\gamma}_{RF}}\Big|\begin{matrix}\varphi_1\\\varphi_2\end{matrix}\right)\right)\left(W\gamma_{th}^{\frac{\xi^2-1}{2}}G_{2,3}^{2,1}\left(\frac{\lambda}{A_0}\sqrt{\frac{\gamma_{th}}{\bar{\gamma}_{FSO}}}\Big|\begin{matrix}1-\xi^2,1\\0,2-\xi^2,-\xi^2\end{matrix}\right)\right)^t. \tag{27}$$

where $\Omega_4=\binom{N-1}{k}\binom{M-1}{t}\binom{t}{u}(-1)^{k+t+u}\frac{N}{k+1}$.

## B. Bit Error Rate

$$\tag{28}$$

Substituting (27) into (19) BER of DBPSK modulation in Negative Exponential atmospheric turbulence with the effect of pointing error becomes equal to:

$$P_e=\frac{1}{2}\int_0^\infty e^{-\gamma}\left\{1-\sum_{k=0}^{N-1}\sum_{t=0}^{M-1}\sum_{u=0}^{t}\Omega_4\,e^{-\frac{(k+u+1)\gamma}{\bar{\gamma}_{RF}}}\left(1-W(\gamma c)^{\frac{\xi^2-1}{2}}\frac{2^{1-\xi^2}}{\sqrt{\pi}}\left(\frac{k+1}{\bar{\gamma}_{RF}}\right)^{\frac{1-\xi^2}{2}}\times\right.\right.$$

$$\left.\left. G_{4,7}^{5,2}\left(\frac{\lambda^2 c(k+1)\gamma}{4A_0^2\bar{\gamma}_{FSO}\bar{\gamma}_{RF}}\Big|\begin{matrix}\varphi_1\\\varphi_2\end{matrix}\right)\right)\left(W\gamma^{\frac{\xi^2-1}{2}}G_{2,3}^{2,1}\left(\frac{\lambda}{A_0}\sqrt{\frac{\gamma}{\bar{\gamma}_{FSO}}}\Big|\begin{matrix}1-\xi^2,1\\0,2-\xi^2,-\xi^2\end{matrix}\right)\right)^t\right\}d\gamma. $$

$$\tag{29}$$

When $t>1$, because of multiplication of three Meijer-G and one exponential functions, the above integral is unsolvable. In the following, new exact expression is derived for BER of the proposed structure in Negative Exponential atmospheric turbulence with the effect of pointing error.

Substituting (39) from Appendix B into (28) leads to a new form of Outage Probability for the proposed structure; substituting the result into (19) and using [37, Eq. 07.34.21.0088.01], exact BER of DBPSK modulation in Negative Exponenial atmospheric turbulence with the effect of pointing error becomes equal to:

$$P_e = \frac{1}{2}\left\{1 - \sum_{k=0}^{N-1}\sum_{t=0}^{M-1}\sum_{u=0}^{t}\sum_{k1=0}^{t}\sum_{n=0}^{\infty}\Omega_4\,F_0^{t-k1}E_n^{(k_1)}\left(\frac{\Gamma(H+1)}{\left(1+\frac{k+u+1}{\bar{\gamma}_{RF}}\right)^{1+H}} - \right.\right.$$

$$\left.\left.\frac{2^{1-\xi^2}}{\sqrt{\pi}}\,W\left(\frac{\bar{\gamma}_{RF}}{k+1}\right)^{\frac{1-\xi^2}{2}}\frac{c^{\xi^2-1}}{\left(1+\frac{k+u+1}{\bar{\gamma}_{RF}}\right)^{H}}G_{5,7}^{5,3}\left(\frac{\lambda^2 c(k+1)}{4A_0^2\bar{\gamma}_{FSO}(\bar{\gamma}_{RF}+k+u+1)}\left|\begin{matrix}-H,\varphi_1\\\varphi_2\end{matrix}\right.\right)\right)\right\}.$$

$$(30)$$

## V. Simulation Results

In this section, analytical and simulation results of performance evaluation of the proposed structure are compared. In MATLAB simulations, Negative Exponential random variable can be generated using *exprnd(.)* command. Radial displacement of pointing error (which is assumed zero mean), has Rayleigh distribution and can be generated by complex addition of two Gaussian random variables using *randn(.)* command. Number of transmitted bits are $10^6$. After generating DBPSK data samples, other parts of the simulation including data transmission, reception, amplification or demodulation, and selection are done exactly the same as what was described in section II (see Fig. 1).

Effects of number of users ($N$) and number of relays ($M$) on the performance of the proposed structure are investigated for both cases of known and unknown CSI. Atmospheric turbulence of FSO link in saturate regime is modeled by Negative Exponential distribution. In order to get closer to the actual results, effect of pointing error is also considered. Fading of RF link has Rayleigh distribution. For simplicity and without loss of generality, FSO and RF links are assumed to have equal average SNR ($\bar{\gamma}_{FSO} = \bar{\gamma}_{RF} = \gamma_{avg}$). In addition, it is assumed that $\eta = 1$, $A_0 = 1$, and $C = 1$. Outage threshold SNR of the proposed system is represented by $\gamma_{th}$.

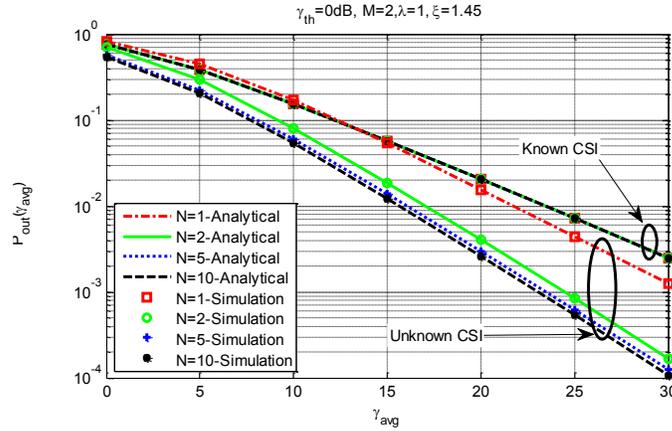

Fig. 1. Outage Probability of the proposed structure as a function of average SNR for various number of users, and both cases of known CSI and unknown CSI, at Negative Exponential atmospheric turbulence with the effect of pointing error with unit variance, when number of relay is $M = 2$, $\xi = 1.45$, and $\gamma_{th} = 10dB$.

In Fig. 1, Outage Probability of the proposed structure is plotted as a function of average SNR for various number of users, and both cases of known CSI and unknown CSI, at Negative Exponential atmospheric turbulence with the effect of pointing error with unit variance, when number of relay is $M = 2$, $\xi = 1.45$, and $\gamma_{th} = 10dB$. As can be seen, in the case of known CSI, performance of the proposed structure is independent of the cell population and in the case of unknown CSI, the proposed structure has slight dependence on the number of users within the cell. At large number of users within the cell, it is more likely to find a signal with desirable $\gamma_{avg}$, therefore the system can perform better. Also the observed outperform of unknown CSI can be seen here. This is because in unknown CSI scheme, the amplification gain is adjusted manually, thereby it can perform better with the cost of more power consumption. In spite of more power consumption of unknown CSI scheme, it has less complexity and processing latency, therefore is especially suitable for communication systems with timing or performance demand.

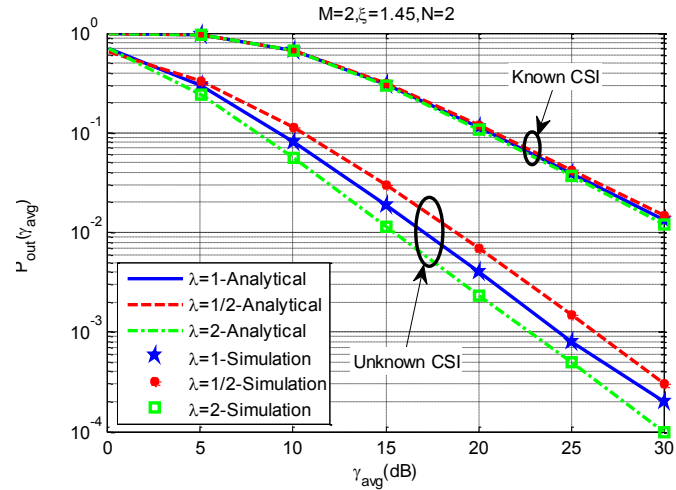

Fig. 2. Outage Probability of the proposed structure as a function of average SNR for various variances of Negative Exponential atmospheric turbulence with the effect of pointing error, and both cases of known CSI and unknown CSI, when number of relays is $M = 2$ and number of users is $N = 2$, and $\xi = 1.45$.

In Fig. 2, Outage Probability of the proposed structure is plotted as a function of average SNR for various variances of Negative Exponential atmospheric turbulence with the effect of pointing error, and both cases of known CSI and unknown CSI, when number of relays is $M = 2$ and number of users is $N = 2$, and $\xi = 1.45$. As can be seen, case of unknown CSI performs better than known CSI. This is because when CSI is unknown, the amplification gain is fixed, usually operators define this gain according to the worst-case scenario and that is exactly why this scheme performs better than known CSI scheme.

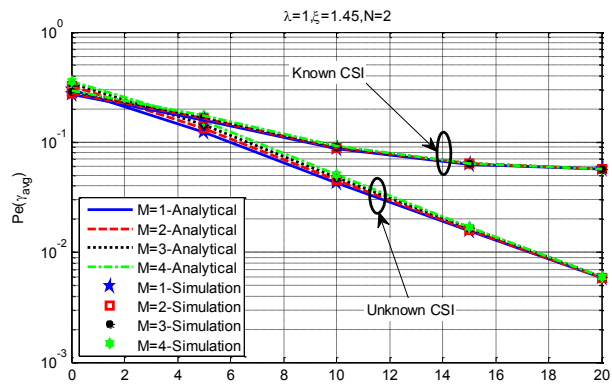

Fig. 3. Bit Error Rate of the proposed structure as a function of average SNR for various number of relays, and both cases of known CSI and unknown CSI, at Negative Exponential atmospheric turbulence with the effect of pointing error with unit variance, when number of users is $N = 2$, and $\xi = 1.45$.

In Fig. 3, Bit Error Rate of the proposed structure is plotted as a function of average SNR for various number of relays, and both cases of known CSI and unknown CSI, at Negative Exponential atmospheric turbulence with the effect of pointing error with unit variance, when number of users is $N = 2$, and $\xi = 1.45$. As can be seen, in both cases of known CSI and unknown CSI, at low $\gamma_{avg}$, there is slight performance difference between different number of relays, and this difference is a bit more in the case of unknown CSI. In both cases of known CSI and unknown CSI, at high $\gamma_{avg}$, system performance is independent of number of relays. Generally, series relaying structure degrades system performance, because of frequent decisions made by relays.

## VI. Conclusion

In this paper, a novel multi-hop hybrid FSO / RF communication system is presented. In this system, an access point connects RF users to the source Base Station via a series hybrid FSO / RF link, and a multi-hop parallel hybrid FSO / RF link connects source and destination Base Stations. This structure is suitable for areas that direct RF connection between mobile users and source Base Station is not possible. Atmospheric turbulence of FSO link in saturate regime is modeled by Negative Exponential distribution. In order to get closer to the actual results, effect of pointing error is also considered. Fading of RF link has Rayleigh distribution. The proposed structure is investigated at various number of users and relays for both cases of known CSI and unknown CSI. Results indicate that the proposed system has low dependence on the number of users; therefore, it is suitable for areas in which population changes frequently. In simulation results, unknown CSI outperforms known CSI, this is because when CSI is unknown, the amplification gain is fixed and chosen manually. In this paper, this gain is defined according to the worst-case scenario and that is exactly why this scheme performs better than the known CSI scheme.

## Appendix A

### The CDF of Negative Exponential atmospheric turbulence with the effect of pointing error

The marginal probability density function (pdf) of pointing error and Negative Exponential atmospheric turbulence are respectively as follows [8]:

$$f_{h_p}(h_p) = \frac{\xi^2}{A_0^{\xi^2}} h_p^{\xi^2-1}; 0 \leq h_p \leq A_0 , \tag{31}$$

$$f_{h_a}(h_a) = \lambda e^{-\lambda h_a} \tag{32}$$

where $1/\lambda^2$ is Negative Exponential atmospheric turbulence variance, $A_0 = [\text{erf}(v)]^2$, $v = \sqrt{\pi/2}\, R/w_b$, R is the radius of the Receiver aperture, $w_b$ is the received beam size. The joint pdf of Negative Exponential atmospheric turbulence with the effect of pointing error is as follows:

$$\tag{33}$$

$$f_h(h) = \int_{\frac{h}{A_0}}^{\infty} f_{h_a}\left(\frac{h}{h_a}\Big| h_a\right) f_{h_a}(h_a) dh_a = \int_{\frac{h}{A_0}}^{\infty} \frac{\xi^2}{A_0^{\xi^2}} \left(\frac{h}{h_a}\right)^{\xi^2-1} \lambda e^{-\lambda h_a} dh_a$$

$$= \frac{\xi^2 \lambda^{\xi^2-1}}{A_0^{\xi^2}} h^{\xi^2-1} \Gamma\left(2 - \xi^2, \frac{\lambda h}{A_0}\right).$$

According that $\gamma = \bar{\gamma}_{FSO} h^2$, and using Equivalent Meijer-G form of $\Gamma\left(2 - \xi^2, \frac{\lambda}{A_0}\sqrt{\frac{\gamma}{\bar{\gamma}_{FSO}}}\right)$ as $G_{1,2}^{2,0}\left(\frac{\lambda}{A_0}\sqrt{\frac{\gamma}{\bar{\gamma}_{FSO}}}\Big| \begin{matrix} 1 - \xi^2, 1 \\ 0, 2 - \xi^2, -\xi^2 \end{matrix}\right)$, the pdf of Negative Exponential atmospheric turbulence with the effect of pointing error becomes equal to:

$$f_{\gamma_{FSO}}(\gamma) = W \gamma^{\frac{\xi^2}{2}-1} G_{1,2}^{2,0}\left(\frac{\lambda}{A_0}\sqrt{\frac{\gamma}{\bar{\gamma}_{FSO}}}\Big| \begin{matrix} 1 \\ 0, 2 - \xi^2 \end{matrix}\right), \tag{34}$$

where $G_{p,q}^{m,n}\left(z\Big| \begin{matrix} a_1, a_2, ..., a_p \\ b_1, b_2, ..., b_q \end{matrix}\right)$ is Meijer-G function [37, Eq. 07.34.02.0001.01], $\xi^2 = \omega_{Z_{eq}}/(2\sigma_s)$ is the ratio of the equivalent received beam radius ($\omega_{Z_{eq}}/2$) to the standard deviation of pointing errors at the receiver ($\sigma_s$), $W = \frac{\xi^2 \lambda^{\xi^2-1}}{2A_0^{\xi^2} \bar{\gamma}_{FSO}^{\xi^2/2}}$. Integrating (46), the CDF of Negative Exponential atmospheric turbulence with the effect of poiting error becomes as follows:

$$F_{\gamma_{FSO}}(\gamma) = W\gamma^{\frac{\xi^2-1}{2}} G_{2,3}^{2,1}\left(\frac{\lambda}{A_0}\sqrt{\frac{\gamma}{\bar{\gamma}_{FSO}}}\,\bigg|\,\begin{matrix} 1-\xi^2, 1 \\ 0, 2-\xi^2, -\xi^2 \end{matrix}\right).$$ 

(35)

## Appendix B

### EXACT CDF OF NEGATIVE EXPONENTIAL ATMOSPHERIC TURBULENCE WITH THE EFFECT OF POINTING ERROR

Using [37, Eq.07.34.26.0004.01] , the CDF of Negative Exponential atmospheric turbulence with the effect of pointing error is equal to:

(36)

$$F_{\gamma_{FSO}}(\gamma) = \frac{\Gamma(2-\xi^2)\Gamma(\xi^2)}{\Gamma(1+\xi^2)}\,{}_2F_2\left(\xi^2, 0; \xi^2-1, \xi^2+1; -\frac{\lambda}{A_0}\sqrt{\frac{\gamma}{\bar{\gamma}_{FSO}}}\right) +$$

$$\frac{\Gamma(\xi^2-2)\Gamma(2)}{\Gamma(\xi^2-1)\Gamma(3)}\left(\frac{\lambda}{A_0}\sqrt{\frac{\gamma}{\bar{\gamma}_{FSO}}}\right)^{2-\xi^2}{}_2F_2\left(2, 2-\xi^2; 3-\xi^2, 3; -\frac{\lambda}{A_0}\sqrt{\frac{\gamma}{\bar{\gamma}_{FSO}}}\right)$$

Using [37, Eq. 07.23.02.0001.01], the above expression becomes equal to:

$$F_{\gamma_{FSO}}(\gamma) = F_0\gamma^{\frac{\xi^2-1}{2}} + \sum_{n=0}^{\infty} E_n\gamma^{\frac{n+1}{2}},$$

(37)

where $F_0 = \frac{\xi^2\lambda^{\xi^2-1}}{2A_0^{\xi^2}\bar{\gamma}_{FSO}^{\xi^2/2}}\frac{\Gamma(2-\xi^2)\Gamma(\xi^2)}{\Gamma(1+\xi^2)}$, and $E_n = \frac{\xi^2}{2}\frac{\Gamma(\xi^2-2)\Gamma(2)}{\Gamma(\xi^2-1)\Gamma(3)}\frac{(2)_n(2-\xi^2)_n(-1)^n}{(3)_n(3-\xi^2)_n n!}\frac{\lambda^{n+1}}{A_0^{n+2}}\frac{1}{\bar{\gamma}_{FSO}^{\frac{n+2}{2}}}$.

Using (57), expression $\left(W\gamma_{th}^{\frac{\xi^2-1}{2}} G_{2,3}^{2,1}\left(\frac{\lambda}{A_0}\sqrt{\frac{\gamma_{th}}{\bar{\gamma}_{FSO}}}\,\bigg|\,\begin{matrix} 1-\xi^2, 1 \\ 0, 2-\xi^2, -\xi^2 \end{matrix}\right)\right)^t$ in (21) can be expanded by trinomial expansion as:

(38)

$$\sum_{k_1=0}^{t}\binom{t}{k_1}(F_0\gamma_{th})^{t-k_1}\left(\sum_{n=0}^{\infty} E_n\gamma_{th}^{\frac{n+1}{2}}\right)^{k_1},$$

after some mathematical simplification, it becomes as:

$$\sum_{k_1=0}^{t} \sum_{n=0}^{\infty} \binom{t}{k_1} F_0^{t-k_1} E_n^{(k_1)} \gamma_{\text{th}}^H,$$

(39)

where $H = \frac{n+\xi^2+k_1+(t-k_1)(\xi^2-1)}{2}$.

<h2 style="text-align:center">REFERENCE</h2>